\documentclass[preprint,12pt]{elsarticle}

\usepackage{amssymb}
\usepackage{url}
\usepackage{amsmath}
\usepackage{xcolor}
\usepackage{soul}
\newcounter{bla}

\begin{document}

\begin{frontmatter}



\title{JuTrack: a Julia package for auto-differentiable accelerator modeling and particle tracking}


\author[a]{Jinyu Wan\corref{author}}
\ead{wan@frib.msu.edu}
\author[a]{Helena Alamprese}
\author[a]{Christian Ratcliff}
\author[b]{Ji Qiang}
\author[a]{Yue Hao\corref{author}}
\ead{haoy@frib.msu.edu}

\cortext[author] {Corresponding author.}
\address[a]{Facility for Rare Isotope Beams, Michigan State University, East Lansing, 48824, USA}
\address[b]{Lawrence Berkeley National Laboratory, Berkeley, CA 94720, USA}

\begin{abstract}
Efficient accelerator modeling and particle tracking are key for the design and configuration of modern particle accelerators. In this work, we present JuTrack, a nested accelerator modeling package developed in the Julia programming language and enhanced with compiler-level automatic differentiation (AD). With the aid of AD, JuTrack enables rapid derivative calculations in accelerator modeling, facilitating sensitivity analyses and optimization tasks. We demonstrate the effectiveness of AD-derived derivatives through several practical applications, including sensitivity analysis of space-charge-induced emittance growth, nonlinear beam dynamics analysis for a synchrotron light source, and lattice parameter tuning of the future Electron-Ion Collider (EIC). Through the incorporation of automatic differentiation, this package opens up new possibilities for accelerator physicists in beam physics studies and accelerator design optimization.
\end{abstract}

\begin{keyword}
Automatic differentiation \sep Particle accelerator \sep Beam dynamics \sep Julia \sep Design optimization
\end{keyword}

\end{frontmatter}

\section{Introduction}
\label{sec1}
Particle accelerators are critical scientific facilities that generate particle beams and accelerate them to energies spanning from KeV to TeV level \cite{Wilson01}. These massive machines are vital tools in various fields of science and technology, enabling groundbreaking research to explore fundamental particles \cite{Hoffmann07}, develop new materials \cite{Liss16}, advance medical imaging and treatment techniques \cite{Hall22}, and perform a variety of cutting-edge research endeavors. 

The design and optimization of modern particle accelerators require precise modeling of beam dynamics in complex electromagnetic environments, accounting for strong nonlinear interactions within the beam. Over the past few decades, many successful accelerator modeling tools have been developed, such as COSY INFINITY \cite{Berz90}, MAD-X \cite{Grote03}, ELEGANT \cite{Borland00}, AT \cite{Terebilo01}, BMAD \cite{Sagan06}, IMPACT~\cite{qiang2000,qiang2006} and TRACK \cite{Aseev05}. These tools have been instrumental in advancing accelerator physics, providing reliable simulations that have significantly contributed to the design and optimization of numerous accelerator facilities worldwide. Recently, there has been growing interest in developing the digital twin of accelerator facilities \cite{Kafkes21}. The modeling tools will be an essential part of the digital twin for predicting the beam behaviors in the accelerator. Nevertheless, modern accelerators, such as the diffraction-limited storage ring (DLSR) \cite{Eriksson14}, and future colliders, e.g., the Electron-Ion Collider (EIC) \cite{Montag23}, the Future Circular Collider (FCC) \cite{FCC19}, the circular electron positron collider (CEPC) \cite{CEPC19}, and muon colliders \cite{Accettura23}, have become increasingly complex and challenging in the dynamics of the beam. The highly nonlinear beam dynamics in the accelerators strongly limits the dynamic aperture and beam lifetime, making the optimization process extremely computationally intensive.

Auto-differentiation (AD) has emerged as a powerful technique in computational physics and machine learning \cite{Baydin18, Fraysse19, Leng24}, offering notable benefits in sensitivity analysis and optimization tasks. AD enables automatic computation of derivatives, which are essential for gradient-based optimization methods and for analyzing the behavior of physical systems under small perturbations. Applying AD to accelerator modeling has significant potential to improve the efficiency of optimizing complex accelerator systems and provide deeper insights into intricate beam dynamics. For instance, COSY INFINITY and MAD-X can provide derivatives in beam dynamics simulation with truncated power series algebra (TPSA) {\cite{Berz88}}. In recent years, several new software packages that incorporate AD capabilities have been developed. One such innovative package is Cheetah \cite{Kaiser24}, a PyTorch-based differentiable linear beam dynamics code. Another notable example is BMAD, a successful accelerator modeling tool in use since the mid 1990s, which is also planning to transition to Julia with AD capabilities to further enhance its computational performance \cite{Sagan24}. In addition, MAD-NG {\cite{manosperti23}} incorporates generalized truncated power series algebra (GTPSA) {\cite{deniau15}}, enabling it to compute derivatives of arbitrary order.

In this work, we present a new AD-enabled particle tracking package, JuTrack. This package is written in Julia \cite{Bezanson17}, a high-level, high-performance programming language specifically designed for scientific computing. With the aid of Enzyme \cite{Moses20}, an AD plugin working at the compiler level, JuTrack allows automatic and rapid calculation of the derivatives of functions. In addition to standard particle tracking, this package also supports TPSA tracking, where the new particle coordinates are represented by a set of polynomials of initial coordinates. This package offers optional features for simulating collective effects, including transverse space-charge effects, wakefield effects, and beam-beam interactions. 

This paper is organized as follows. In Section 2, we introduce the core features and architecture of JuTrack. Section 3 demonstrates the correctness and performance of JuTrack through a series of benchmark tests. In Section 4, we explore practical applications of JuTrack, including sensitivity analysis of space-charge-induced emittance growth, nonlinear beam dynamics studies, and design optimization tasks. Section 5 is the conclusion of this work.

\section{Design of the package}
\label{sec2}
\subsection{Package overview}
JuTrack is developed in pure Julia. We chose Julia for the development of JuTrack for several compelling reasons. Julia's syntax is easy to read and write, similar to that of high-level languages like Python, which accelerates development time. Additionally, Julia's performance is on par with traditional low-level languages such as C and Fortran due to its Just-In-Time (JIT) compilation using the LLVM framework \citep{Bezanson17, Lattner04}. 

\begin{figure}[!htb]%% placement specifier
%% Use \includegraphics command to insert graphic files. Place graphics files in 
%% working directory.
\centering%% For centre alignment of image.
\includegraphics*[width=.75\columnwidth]{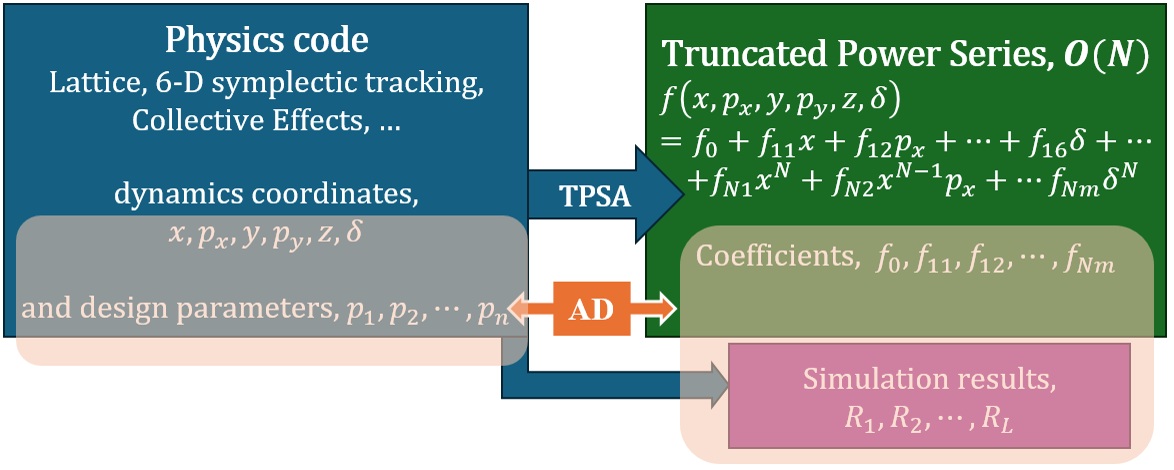}
%% Use \caption command for figure caption and label.
\caption{Schematic diagram of the JuTrack package.}\label{fig1}
%% https://en.wikibooks.org/wiki/LaTeX/Importing_Graphics#Importing_external_graphics
\end{figure}

Figure \ref{fig1} illustrates the functionalities of the package, which consists of two main components: the physics code and the TPSA functions. The primary goal of this package is to ensure that every target physics quantity such as beam emittance and every TPSA coefficient obtained from physics simulation is differentiable with respect to parameters of interest, which typically include the design parameters of the accelerator lattice and the initial coordinates of particles.

\subsection{Implementation of automatic differentiation}
AD is a set of techniques for evaluating the derivatives of functions specified by computer programs. Unlike symbolic differentiation, which manipulates mathematical expressions to find derivatives, or numerical differentiation, which approximates derivatives using finite differences, AD systematically breaks down the function into a sequence of elementary operations. By applying the chain rule of calculus to each elementary operation, AD computes exact derivatives, ensuring high precision and avoiding being deep in the expression hell. 

There are several mature packages with AD capability that have been developed in the accelerator physics community, such as BMAD-X {\cite{Gonzalez23}}, Cheetah, and MAD-NG. BMAD-X is a Python package with library-agnostic tracking routines based on BMAD that allows high-dimensional gradient-based optimization, model calibration, and phase space reconstruction. Cheetah, built on PyTorch {\cite{Paszke19}}, integrates seamlessly with machine learning workflows, enabling advanced applications such as modular surrogate modeling and reinforcement learning. These Python-based packages are more accessible to a wider user base than traditional packages developed in non-scripting languages. Machine learning packges like Pytorch usually supports backward AD. It will be challenging to manage computing graphs for backward AD in complicated simulation code. MAD-NG, which utilizes GTPSA, provides flexibility with arbitrary-order differentiation. Considering in practical optimization tasks, people are usually more concerned with the first-order derivatives, we focus on the first-order derivative computation in the development of our package.

To implement AD to the first order with high efficiency, we utilize a third-party plugin, Enzyme \citep{Moses20}. Enzyme is a high-performance AD compiler plugin for the LLVM compiler framework with the capability of both forward and backward AD. It is capable of calculating derivatives of statically analyzable programs expressed in the LLVM intermediate representation (IR). This compiler plugin is available in any language whose compiler targets LLVM IR, including C, C++, Fortran, Julia, Rust, and others. Unlike conventional source-to-source and operator-overloading tools, Enzyme implements AD at the compiler level, resulting in significantly better computational efficiency \citep{Moses21}. Although Enzyme supports high-order derivatives computation, we only focus on first-order derivatives in this package for best efficiency. The users can extend it to higher-order if necessary.

\subsection{Implementation of particle tracking}
JuTrack supports 6-D tracking of particle's coordinates, $x$, $p_x$, $y$, $p_y$, $z$, and $\delta$. $x$ and $y$ are the transverse coordinates, $p_x$ and $p_y$ are the transverse momentum, $z=-\beta c\tau$ is the path lengthening with respect to the reference particle with $\tau$ as the relative arrival time at location $s$, and $\delta=dp/p_0$ is the energy deviation with respect to the reference particle.

Several common magnet elements, such as dipole, quadrupole, sextupole, octupole and solenoid, as well as radio frequency (RF) cavities for acceleration and deflection have been modeled in JuTrack. The tracking maps for these components have been established. Similar to other symplectic tracking codes, each magnet element in JuTrack can be divided into several slices. To conserve phase space area for long-term tracking, fourth-order symplectic integration is implemented in the tracking process. Within each slice, particles travel according to the symplectic map \citep{Yoshida90}, 
\begin{equation}
    H(p,q)=H_1(p)+H_2(q)
\end{equation}
\begin{equation}
    e^{-:H:}=e^{:-c_1LH_1:}e^{:-d_1LH_2:}e^{:-c_2LH_1:}e^{:-d_2LH_2:}e^{:-c_2LH_1:}e^{:-d_1LH_2:}e^{:-c_1LH_1:}+O(L^5)
\end{equation}
where $L$ is the slice length, and $H$ is the Hamiltonian. $c_1=\frac{1}{2(2-2^{1/3})}$, $c_2=\frac{1-2^{1/3}}{2(2-2^{1/3})}$, $d_1=\frac{1}{2-2^{1/3}}$, and $d_2=-\frac{2^{1/3}}{2-2^{1/3}}$.

Parallel computing is available for multi-particle tracking. By distributing workloads across multiple threads using Julia's built-in support for multi-threading, we can take full advantage of modern multi-core processors to improve the efficiency and scalability of our particle tracking code.

\subsection{Multivariate TPSA module}
The TPSA, first introduced by Berz \citep{Berz88}, has been used in accelerator modeling code since 1990s \citep{Yan94}. A TPSA variable can be represented as a vector of Taylor series coefficients. The operations on the TPSA variables, such as addition and multiplication, are evaluated as vector operations following the rules of TPSA. The single-variable TPSA can be extended to the multivariate form \citep{Chao02}. Taking 6-D variables in our particle tracking regime as an example, the 6-D coordinates can be represented as,
\begin{equation}
\begin{aligned}
    x &= (x_0, 1, 0, 0, 0, 0, 0, \ldots, 0) \\
    p_x &= (p_{x_0}, 0, 1, 0, 0, 0, 0, \ldots, 0) \\
    y &= (y_0, 0, 0, 1, 0, 0, 0, \ldots, 0) \\
    p_y &= (p_{y_0}, 0, 0, 0, 1, 0, 0, \ldots, 0) \\
    z &= (z_0, 0, 0, 0, 0, 1, 0, \ldots, 0) \\
    \delta &= (\delta_0, 0, 0, 0, 0, 0, 1, \ldots, 0)
\end{aligned}
\end{equation}

Multivariate TPSA is widely used for transfer matrix calculation and high-order transfer map analysis for beam dynamics. To implement multivariate TPSA, we have developed overloaded operators for a wide range of functions, including addition, subtraction, multiplication, division, trigonometric, inverse trigonometric, hyperbolic, square root, power, exponential, and logarithmic functions. These operators cover all types of operations required in our particle tracking code. By using these established TPSA variables and overloaded operators, standard particle tracking can be seamlessly replaced by tracking TPSA variables.

Considering the 6-D TPSA coordinates [$x$, $p_x$, $y$, $p_y$, $z$, $\delta$] at a starting location in the accelerator, we can calculate the coordinates after one turn. The first-order coefficients of these final coordinates form the linear transfer matrix, which is fundamental for the calculation of the optics functions, such as the $\beta$ functions and dispersion functions. In addition to TPSA, a finite difference approximation approach is also provided in this package for fast calculation of the first-order transfer map. The derivatives of the transfer map coefficients with respect to specific parameter, such as magnet strengths and initial coordinates of the particle, can be obtained with Enzyme.

\section{Benchmark and validation}
To validate the accuracy and performance of JuTrack, we conduct a series of benchmark tests and comparisons with established accelerator modeling tools. Our primary focus is on particle tracking and optics calculations, essential components for assessing the functionality of our code.

\subsection{Validation of accuracy}
The future Electron-Ion Collider (EIC) \cite{Montag23} is designed to collide spin-polarized beams of electrons and ions to study the properties of nuclear matter in detail via deep inelastic scattering. The Electron Storage Ring (ESR) of the EIC consists of thousands of magnet and cavity components, with a circumference of up to 3.8 km, providing a sufficiently complex system to thoroughly test our package.

\begin{figure}[!htb]%% placement specifier
%% Use \includegraphics command to insert graphic files. Place graphics files in 
%% working directory.
\centering%% For centre alignment of image.
\includegraphics*[width=0.9\columnwidth]{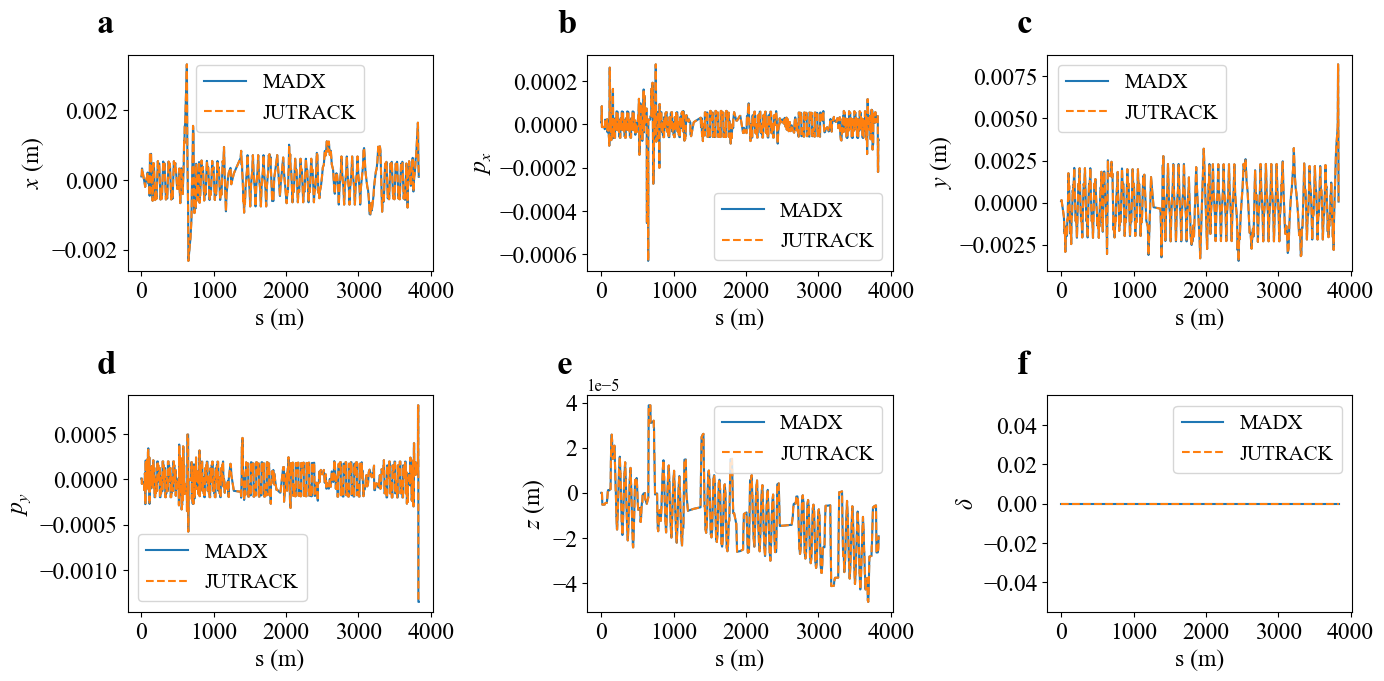}
%% Use \caption command for figure caption and label.
\caption{Comparison of single-particle tracking trajectories within the ESR obtained with JuTrack and MAD-X. (a)-(f) represent the comparisons for $x$, $p_x$, $y$, $p_y$, $z$, and $\delta$, respectively.}\label{fig2}
%% https://en.wikibooks.org/wiki/LaTeX/Importing_Graphics#Importing_external_graphics
\end{figure}

To evaluate the correctness of the particle tracking module in JuTrack, the 6-D trajectory of an electron with random initial transverse coordinates [$x$, $y$] is tracked within the ESR. The results are compared with those of the well-benchmarked Polymorphic Tracking Code (PTC) integration of MAD-X \citep{Schmidt05}, the software used for the EIC design. The comparison is illustrated in Fig. \ref{fig2}. The discrepancy in the final coordinates [$x$, $p_x$, $y$, $p_y$, $z$, $\delta$] between the two codes is close to the round-off error of the computer, which is negligible. 

\begin{figure}[!htb]%% placement specifier
%% Use \includegraphics command to insert graphic files. Place graphics files in 
%% working directory.
\centering%% For centre alignment of image.
\includegraphics*[width=0.8\columnwidth]{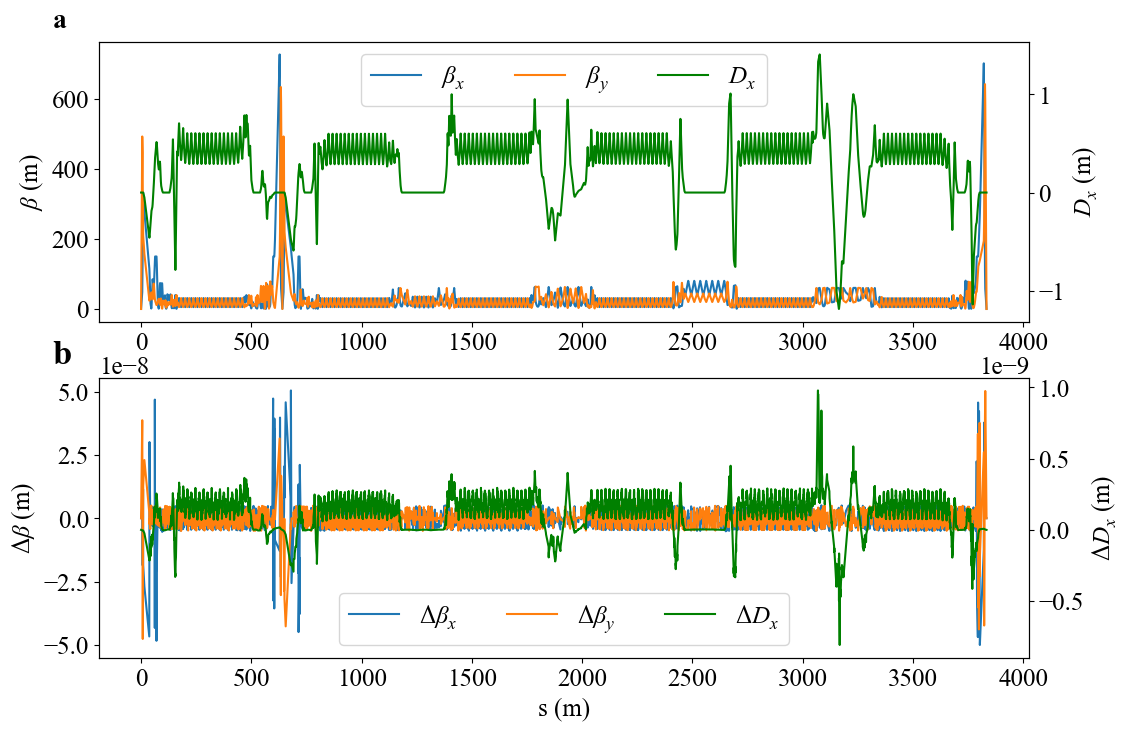}
%% Use \caption command for figure caption and label.
\caption{Illustration of lattice optics for the ESR. (a) $\beta$ functions and dispersion calculated with JuTrack. (b) Difference between the results obtained with JuTrack and MAD-X.}\label{fig3}
%% https://en.wikibooks.org/wiki/LaTeX/Importing_Graphics#Importing_external_graphics
\end{figure}

Using the TPSA tracking method in JuTrack, the transfer matrix of the ESR is obtained. With this transfer matrix, the periodic Twiss parameters of the lattice are solved. Fig. \ref{fig3} presents the comparison of $\beta$ functions and dispersion obtained with JuTrack and MAD-X. The differences in the beta functions and dispersion obtained with the two codes are less than 10$^{-7}$m and 10$^{-9}$m, respectively. The tunes of betatron oscillation are 50.08 and 46.14, which are identical to the results from MAD-X. 

These benchmarks demonstrate that JuTrack provides highly accurate particle tracking and optics calculations, comparable to established tools like MAD-X, validating its effectiveness for accelerator physics research.

\subsection{Performance evaluation}
To evaluate the computational efficiency of JuTrack in handling complex simulations, we conduct tests by tracking 1000 particles with random initial coordinates within the ESR over 1000 turns. To prevent particle loss from affecting the tracking time, the initial coordinates are confined to a small range, significantly smaller than the dynamic aperture, ensuring that no particles are lost during the tracking process.

The tracking test is conducted on a personal laptop equipped with 14 physical cores of 2.6 GHz 13th Gen Intel(R) CPUs. The computation time required by JuTrack and MAD-X is shown in Table 1. For both packages, the same 4th-order symplectic integration is implemented, with each magnet in the accelerator divided into 10 slices.

Without implementing parallel computing, the computational efficiency of multi-particle tracking is of the same order of magnitude for both packages. This implies Julia provides similar or even better efficiency compared to traditional low-level languages. Additionally, JuTrack's parallel computing capability significantly reduces the computation time by an order of magnitude. The efficiency of JuTrack can be further enhanced with more powerful computing devices equipped with additional processors, allowing for even larger and more complex simulations to be performed within practical time frames.

We also evaluate the efficiency of AD and compare it with the traditional numerical approach using finite differences. Our tests include a variety of functions, such as simple arithmetic operations with TPSA, particle tracking simulations, Twiss parameter calculations, and space charge computations. The computational times for these original functions ranged from microseconds to several minutes, depending on their complexity. For all tested functions, the computation time required for AD is almost twice that of the finite difference method, which is primarily due to the overhead associated with AD.

Despite the increased computational time, AD offers significant advantages in terms of accuracy and reliability. Finite difference methods are sensitive to the choice of step size and often struggle with numerical noise caused by truncation errors and floating point precision. In contrast, AD provides exact derivatives up to machine precision without the need to select an appropriate step size. This makes AD particularly advantageous for ill-conditioned functions, which are frequently seen in the nonlinear beam dynamics.

\begin{table}[t] %% placement specifier
\centering %% For centre alignment of tabular.
\caption{Comparison of computation time.}\label{tab1}
\begin{tabular}{l c} %% Table column specifiers
\hline %% Adds a horizontal line at the top
\textbf{Method} & \textbf{Time (seconds)} \\ %% Header row with centered alignment
\hline %% Adds a horizontal line below the headers
MAD-X PTC & 5420.7 \\
JuTrack single-threading & 3277.5 \\
JuTrack multi-threading & 528.69 \\
\hline %% Adds a horizontal line at the bottom
\end{tabular}
\end{table}

\section{Application to practical problems}
\label{sec3}
In this section, we apply JuTrack to several complex, practical problems in accelerator physics. We demonstrate how the derivatives obtained through AD can be utilized for sensitivity analysis of space charge-induced emittance growth, nonlinear beam dynamics studies, and design optimization.

\subsection{Sensitivity analysis of emittance growth due to space charge}
The nonlinear space charge effects from the Coulomb interaction inside a charged particle beam can significantly impact the beam quality in a high intensity or a high brightness accelerator by causing emittance growth, beam halo, and even particle losses. Following the numerical spectral Galerkin method \cite{Gottlieb77}, the one-step symplectic transfer map of the transverse space charge kick on an individual particle in a rectangular perfectly conducting pipe can be written as \cite{Qiang23},
\begin{equation}
    \begin{aligned}
    p_{xi}(\tau)=p_{xi}(0)-\tau \frac{K}{2} \sum_{l=1}^{N_l} \sum_{m=1}^{N_m} \phi ^{lm}\alpha_l \cos(\alpha_lx_i) \sin(\beta_my_i) \\
    p_{yi}(\tau)=p_{yi}(0)-\tau \frac{K}{2} \sum_{l=1}^{N_l} \sum_{m=1}^{N_m} \phi ^{lm}\beta_m \sin(\alpha_lx_i) \cos(\beta_my_i),
    \end{aligned}
\end{equation}
where $K=qI/2\pi \epsilon_0p_0v_0^2\gamma_0^2$ is the generalized perveance, $\tau$ is the effective length, $i$ is the index of the $i$th particle, $a$ is the horizontal width of the pipe, $b$ is the vertical width of the pipe, $\alpha_l=l\pi/a$ and $\beta_m=m\pi/b$. The space charge potential $\phi ^{lm}$ is given in the spectral domain as,
\begin{equation}
    \phi ^{lm}=4\pi\frac{4}{ab}\frac{1}{N_p}\sum_{j=1}^{N_p}\frac{1}{\gamma^2_{lm}}\sin(\alpha_lx_j)sin(\beta_my_j),
\end{equation}
where $N_p$ is the number of particles and $\gamma^2_{lm}=\alpha_l^2+\beta_m^2$ and $j$ is the index of the $j$th particle.

As discussed in \cite{Qiang23}, the emittance growth due to space charge effects is related to the lattice parameters. In the following, we will study the derivatives of the final emittance of a 1-GeV coasting proton beam passing through a transversely focusing FODO cell with respect to these lattice parameters. The FODO cell consists of two focusing quadrupoles, Q1 and Q2, each with a length of 0.1 m and equal but opposite focusing strengths, and three drift spaces, D1 and D3 of 0.2 m each, and D2 of 0.4 m. The beam has a current of 200 A with a normalized emittance of 1 mm·mrad. The horizontal and vertical aperture sizes of the rectangular pipe are both 13 mm.

\begin{figure}[!htb]%% placement specifier
%% Use \includegraphics command to insert graphic files. Place graphics files in 
%% working directory.
\centering%% For centre alignment of image.
\includegraphics*[width=0.5\columnwidth]{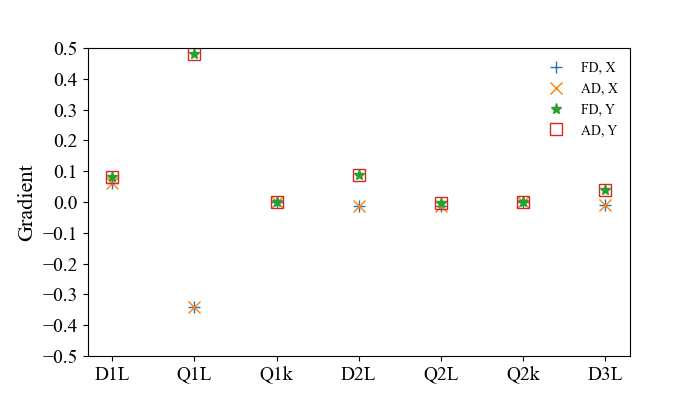}
%% Use \caption command for figure caption and label.
\caption{Derivative of final horizontal and vertical emittance with respect to seven lattice parameters. FD and AD represent the derivatives obtained with finite difference approximation and automatic differentiation, respectively.}\label{fig4}
%% https://en.wikibooks.org/wiki/LaTeX/Importing_Graphics#Importing_external_graphics
\end{figure}

In this example, we perform particle tracking simulations using 5000 macro particles and 12×12 spectral modes. The final emittance (normalized by the initial emittance) is calculated using the JuTrack particle tracking module. In the meantime, the derivatives of the final emittance with respect to the lattice parameters are automatically obtained via AD. Figure \ref{fig4} presents these derivatives with respect to the lengths of the drift spaces (D1L, D2L and D3L), and the lengths (Q1L and Q2L) and strengths (Q1k and Q2k) of the quadrupoles. For validation purposes, we also compute approximate derivatives using the finite difference method. It is found that the results of finite difference approach nearly converge when the difference is less than 10$^{-4}$. However, when the difference is too small, e.g., less than 10$^{-9}$, the round-off errors will dominate. Therefore, we choose a reasonable difference of 10$^{-6}$ in the calculation. The close agreement between the AD results and the finite difference approximations confirms the accuracy of the AD computations. Our results show that the emittance variation is more sensitive to the length of the first quadrupole, which is consistent with the findings reported in \cite{Qiang23}.

\subsection{Application to beam dynamics analysis}
The dynamic aperture (DA) is a fundamental concept in the design and operation of storage ring accelerators. DA represents the volume in phase space where particles can circulate stably without being lost due to nonlinear effects. DA is usually determined through extensive particle tracking simulations, which involve tracking a large number of particles with varying initial conditions over a sufficiently long period in the accelerator. While accurate, this method is computationally expensive, often requiring significant computational resources and time. To fast optimize DA, accelerator physicists proposed an alternative approach, the resonance driving terms (RDTs) \cite{Yang11}. By calculating and minimizing these terms, one can enhance the DA, which is widely used in accelerator design and optimization.

In the normal form analysis, the Hamiltonian of the whole ring is normalized in resonance bases. The Lie generator is written as \cite{Bengtsson97},
\begin{equation}
    h = \sum h_{abcde}h_x^{+a}h_x^{-b}h_y^{+c}h_y^{-d}\delta^e
\end{equation}
where $h^{\pm}_x\equiv\sqrt{2J_x}e^{\pm i\phi_x}=x\mp ip_x$, and $(J, \phi)$ are action-angle variables. The nonlinear driving terms $h$ can be grouped by their order $n=a+b+c+d+e$ and each $h_{abcde}$ is an explicit analytical function of magnet strengths, beta functions, and dispersions.

\begin{figure}[!htb]%% placement specifier
%% Use \includegraphics command to insert graphic files. Place graphics files in 
%% working directory.
\centering%% For centre alignment of image.
\includegraphics*[width=0.7\columnwidth]{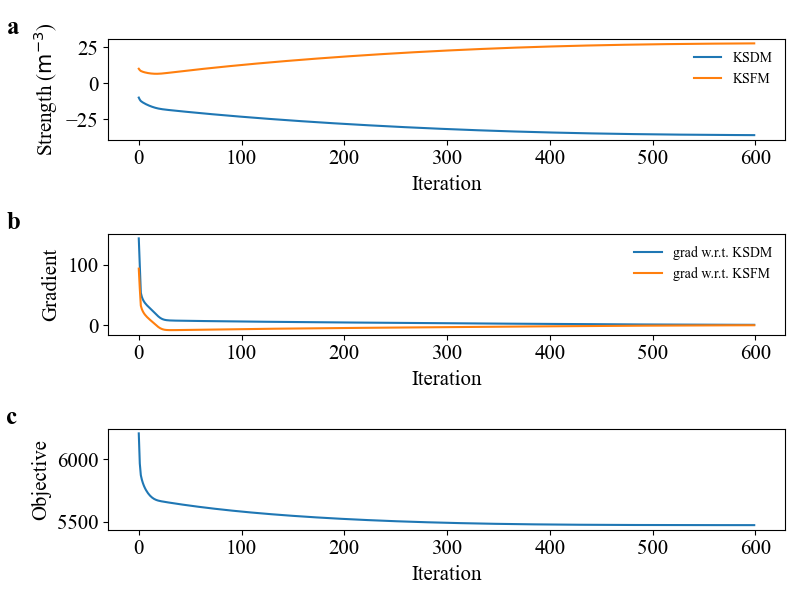}
%% Use \caption command for figure caption and label.
\caption{Optimization process of the RDTs for the SPEAR3 ring. (a) shows the evolution of the strengths of two sextupole families, SDM and SFM, respectively. (b) shows the evolution of the derivatives of the objective function with respect to the sextupole strengths. (c) is the change of the objective function, which is the summation of five geometric RDTs.}\label{fig5}
\end{figure}

As a practical example, we apply JuTrack to the Stanford Positron Electron Asymmetric Ring (SPEAR3) \cite{Corbett1998}, a third-generation synchrotron light source at SLAC National Accelerator Laboratory. Five geometric RDTs, $h_{21000}$, $h_{10110}$, $h_{30000}$, $h_{10200}$, and $h_{10020}$, are calculated for the ring based on the lattice functions including Twiss parameters, dispersion, and phase advances. To demonstrate the effectiveness of the AD-derived derivatives, we select two families of sextupoles installed in the SPEAR3 ring, namely, SDM and SFM, for the optimization of these RDTs. The derivatives of the RDTs with respect to the sextupole strengths can be automatically obtained in the calculation. 

Using the derivatives, a simple gradient-descent optimization is implemented to suppress the geometric RDTs. Figure 5 shows that the RDTs are reduced through the optimization. After approximately 500 iterations, the algorithm converged to a local minimum where the gradients with respect to both SDM and SFM sextupole strengths approached zero. This result indicates that the RDTs can be effectively minimized with JuTrack, which could lead to an improvement in the DA. The calculation of the RDTs and their derivatives takes less than one second on a personal laptop, providing the potential for rapid optimization of the DA.

\subsection{Application to design optimization}
One of our primary goals of implementing AD in accelerator modeling is to optimize various design parameters to achieve specific beam properties, such as minimizing natural emittance, maximizing DA, and improving overall beam stability. With the derivatives obtained from AD, we can simplify complex optimization tasks into gradient-descent problems. In this section, we demonstrate how to apply AD in the design optimization of particle accelerators, using the design of the ESR of the EIC as an example.

\begin{figure}[!htb]%% placement specifier
%% Use \includegraphics command to insert graphic files. Place graphics files in 
%% working directory.
\centering%% For centre alignment of image.
\includegraphics*[width=0.8\columnwidth]{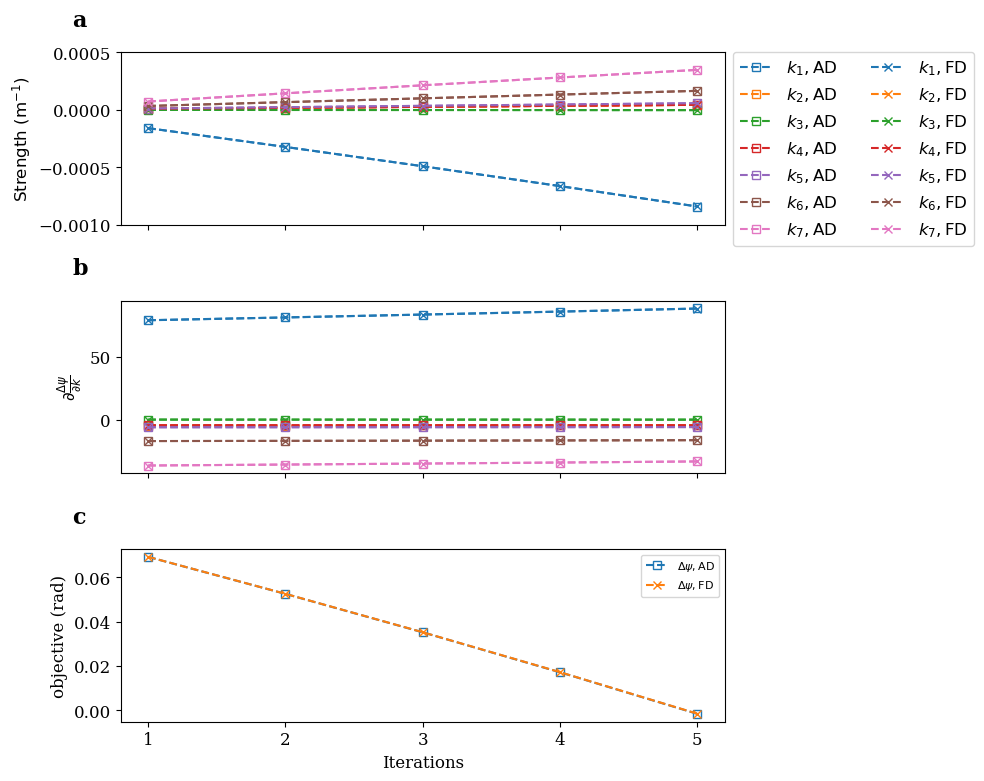}
%% Use \caption command for figure caption and label.
\caption{Tuning the magnet strengths to optimize the phase advance of a crab cavity pair in ESR. AD and FD represent using derivatives obtained from automatic differentiation and finite difference, respectively. (a) shows change of the magnet strengths from $k_1$ to $k_7$. (b) shows change of the derivatives of $\Delta\psi$. (c) shows evolution of the optimization objective, $\Delta\psi$. 2$\pi$ is the target phase advance.}\label{fig6}
%% https://en.wikibooks.org/wiki/LaTeX/Importing_Graphics#Importing_external_graphics
\end{figure}

The electron bunch in the ESR is tilted in the $z-x$ plane by the crab cavities to compensate for the geometric luminosity loss \cite{Xu22}. The phase advance of the crab cavity pair needs to be set correctly to create a local crabbing bump in the interaction region. To calculate the phase advance $\Delta\psi$, we create a function using JuTrack, $f(k_i)=\Delta\psi, i=1,2,...,7$. $k_1$ to $k_7$ represent the strengths of seven quadrupoles installed between the crab cavity pair, which are the tunable parameters to control the $\Delta\psi$. 

With the AD capabilities of JuTrack, the derivatives of $\Delta\psi$ with respect to $k_i$ are calculated during the calculation of $\Delta\psi$. These derivatives are then utilized to perform a gradient-descent optimization, where the strengths $k_1$ to $k_7$ are simultaneously adjusted based on the derivatives with a fixed step size. For comparison, another optimization is conducted using derivatives approximated by the finite difference approach. Figure \ref{fig6} illustrates the evolution of the multivariate optimization process. Within five iterations, the gradient-descent algorithm successfully converges to the optimal solution, achieving a phase advance $\Delta\psi$ of $2\pi$.

The results indicate that the derivatives obtained through AD and finite difference are almost the same. This is because the linear optics involved in the problem do not introduce complicated nonlinear correlations that could lead to ill-conditioned systems. As a result, the finite difference method is sufficiently accurate for this case. However, for more complicated problems, particularly those involving nonlinear forces from elements such as sextupoles, the finite difference approach may struggle due to its limited accuracy. In such scenarios, AD remains a reliable choice to provide precise derivatives.

\section{Conclusion}
\label{sec4}
An AD-enabled particle accelerator modeling package, JuTrack, has been developed using the Julia programming language. With the aid of Enzyme, an LLVM-level plugin for AD calculation, JuTrack can automatically obtain the derivatives of functions in accelerator modeling simulation. Our benchmark tests show that the computational efficiency of particle tracking using JuTrack is comparable to other packages written in traditional low-level languages such as Fortran and C. Furthermore, by distributing workloads across multiple cores using Julia's parallel computing feature, computational time for multi-particle tracking can be further reduced.

We demonstrate that the derivatives provided by JuTrack can be effectively used for sensitivity analysis of space charge-induced emittance growth, nonlinear beam dynamics analysis, and design optimization of the lattice. This capability is crucial for understanding how lattice parameters influence beam stability and performance, enabling more informed decisions in accelerator design and configuration. 

The development and implementation of JuTrack provide accelerator physicists with an efficient, precise, and flexible tool for beam dynamics studies and design optimization. The package is actively being improved, with plans to incorporate more complex models in the future, such as three-dimensional space charge effects and coherent synchrotron radiation effects. These forthcoming features aim to broaden JuTrack's applicability and utility in advanced accelerator research.

\section*{Acknowledgement}
This work is supported by the DOE Office of Science, with award number DE-SC0024170.



%% References


\bibliographystyle{elsarticle-num}






% \end{thebibliography}

\end{document}